**Charles Lewis Brook: third Director of the BAA Variable Star Section**

Jeremy Shears

**Abstract**

Charles Lewis Brook, MA, FRAS, FRMetS (1855 – 1939) served as Director of the BAA Variable Star Section from 1910 to 1921. During this time he was not merely interested in collecting the observations of the members (to which he also contributed), but he also spent considerable amounts of time analysing the data and preparing numerous publications on the findings. This paper discusses Brook's life and work, with a particular focus on his contribution to variable star astronomy.

**Introduction**

Charles Lewis Brook (1855 – 1939; Figures 1 and 2) was the third Director of the BAA Variable Star Section (VSS) serving from 1910 to 1921. As well as being an enthusiastic observer of variable stars, he undertook the analysis of variable star observations submitted by others, writing many papers and memoirs on the subject. His astronomical interests were not confined to variables and he observed a range of other astronomical objects. In addition to astronomy, he was an active meteorologist. He led a busy professional life as a Director of the thread and textile company Messrs. J. & P. Coates Ltd. He spent most of his life at Meltham, near Huddersfield, in Yorkshire (1). Brook was a member of the BAA, a Fellow of the Royal Astronomical Society and a Fellow of the Royal Meteorological Society. This paper discusses Brook's life and work, with a particular focus on his contribution to variable star astronomy. It draws on four separate obituaries: one written for the BAA (2) by his successor as VSS Director, W.M. Lindley (1891 - 1972), one written for the Royal Astronomical Society by Annie (A.S.D.) Maunder (1868-1947) (3), one for the Royal Meteorological Society (4) and a brief one published in the Huddersfield Examiner (5). In addition I have reviewed Brook's published work, especially the papers he published in the BAA Journal.

**Brook's personal life and professional activities**

Charles Lewis Brook was born on 12 June 1855 at Meltham, Yorkshire, where he lived for most of his life, except whilst at school and University (6). His father was Charles John Brook (1829-1857), who died when Brook was an infant, and his mother was Mary Brook (née Jones; 1835-1919). Mary was the daughter of Rev. Lewis Jones (7) and it was from his maternal grandfather that he gained his middle name of Lewis. Brook was usually referred to by the name of Lewis, rather than Charles, to avoid confusion as there were several other members of the Brook family with Charles as their Christian name (8). Brook had two sisters: Ruth Mary Brook (1856-1932; Figure 3), who shared his interest in astronomy, and Esther Frost Brook (1857-1944). Brook was sent as a boarder to Uppingham School (9) in Rutland, where he excelled at sport (10). He earned his Cricket colours twice (1872 and 1873), representing Uppingham School in their match against Surrey Club on 30 June and 1 July 1873 at the Oval (Surrey won by 9 wickets), and his Rugby colours three times (1871, 1872 and 1873) (11). In later life he became a trustee of Uppingham School. He left Uppingham in May 1874 and went up to Trinity College Oxford where he took an M.A. degree.

In adulthood, Brook lived at Harewood Lodge in Meltham (12), remaining a bachelor. Meltham has a long association with the Brook family, which had established the thread manufacturer, Jonas Brook & Brothers (the brothers being James and Joseph). In its heyday at the end of the 19$^{th}$ century the mill complex at Meltham Mills employed over 1000 workers. The family were well-known philanthropists who took care of their employees, building housing and a convalescence home for them, as well as providing them with pensions. The Brook family built Meltham Hall, St. James' Church in Meltham Mills (Figure 4) and Christ Church at Helme (Figure 5), a village near Meltham. Brook himself was an active member of St. James' Church. His interest in Helme Church, which was built in memory of his father, lasted from the time when he laid the foundation stone at the tender age of three (13), to the offer of a bequest in his will (4) .

In 1896 Jonas Brook & Bros. amalgamated with three other thread manufacturers, J. & P. Coates Ltd. of Paisley, Renfrewshire, James Chadwick & Bros. Ltd of Bolton, and Clark & Co. of Paisley to form an enlarged firm of J. & P. Coates Ltd (14) (15) (16), and which later became known as United Thread Mills Ltd. Each company maintained its identity with the preservation of names and trademarks. The company logo for Jonas Brook & Bros was a goat's head (Figure 6) and this adorned their products which were sold worldwide (17). At the time of the amalgamation, the market value of the combined company was £22 million. The headquarters were in Glasgow, with 17 manufacturing centres and 21,000 employees around the world, the UK workforce totalling around 11,000. Some members of the Brook family, who were also Directors, migrated to Scotland as a result of the merger (8). The company continued trading as J. & P. Coats Ltd. through the first half of the 20$^{th}$ century; several amalgamations followed leading to the formation of Coats Viyella PLC, which eventually became Coats PLC in 2001 (18).

Brook himself was a Director of J. & P. Coats and was clearly a man of considerable personal wealth. He continued his family's tradition of charitable support and was active in both diocesan and hospital work (19). And, as we shall see, he made significant contributions to the publication of BAA and RAS memoirs. He was an enthusiastic supporter of the Meltham Mills Cricket Club (20) and the Huddersfield chess club (4).

Brook died on 9 May 1939, having been ill for some years (5). He was buried in the churchyard of St. James' Church, Meltham Mills on 11 May (21) and his gravestone is shown in Figure 7. In 1949, the lighting in Helme Church was installed in memory of Brook, his sister Ruth Mary, and her husband, Arthur Charles Brook (Figure 8).

**BAA Variable Star Section**

Brook was elected a Fellow of the Royal Astronomical Society in January 1899, his application form being signed by the Rev. T.H.E.C. Espin (1858-1934) "from personal knowledge" (22). Espin was one of the most accomplished amateur astronomers of the period (23) (24). Brook's RAS application and Fellowship forms are shown in Figures 9 and 10. In June of the same year Brook was elected as a member of the BAA (25) (26). He frequently attended meetings of the Association in London, where he also spoke from time to time and he served as Vice President 6 times between 1918 and 1926 (27). How soon he became interested in variable stars is not known, but by January 1902 he had began submitting observations to the BAA Variable Star Section. We know he also observed Nova Persei 1901, as will be discussed later; and one wonders to what extent the appearance of this bright nova might have stimulated his interest in variable stars. The VSS database contains a total of 7648 of his observations, although this is likely to be an underestimate as not all observations have yet been entered, especially those from the early part of the 20$^{th}$ century (28). Most of his observations were made with a 9 inch (22.9 cm) reflector at his observatory at Harewood Lodge (29). His last observations were made in 1924 when poor health forced him to stop (2).

Brook's grasp of variable star science must have become evident fairly early on (30), for in 1908 he edited the RAS Memoir on Norman Pogson's (1829-1891) observations of 31 variable stars (31). Pogson had made his observations both from England and, from 1861, whilst government astronomer at the Madras Observatory in India. In the introduction to the Memoir, Prof. H.H. Turner (32) (1861-1930; Figure 11), Savilian Professor of Astronomy at Oxford, drew generous tribute to Brook's work in analysing Pogson's observations and preparing them for publication. In addition, Brook contributed one-third of the cost of printing the memoir. H.H. Turner was later to provide advice to Brook on variable star science when Brook became VSS Director (33).

The BAA VSS was founded in October 1890 and the first Director was the Dublin-based amateur astronomer John Ellard Gore (1845-1910), who was a well known author of popular astronomy books as well as being an active variable star observer and a computer of binary system elements., However, the Section struggled to make headway and towards the end of Gore's tenure membership was shrinking (24). Gore was succeeded by Ernest Elliott Markwick (1853-1925; Figure 12) as Director in 1899. Markwick was a natural leader and soon set about re-invigorating the Section, focussing observers' attention on a small programme of core stars and publishing regular reports in the *Journal* as well as *Memoirs* (24).The Section was thus in good health when Markwick handed the Director's baton to Brook in January 1910.

Brook introduced himself as the new VSS Director in a letter to the JBAA written on 15 January 1910 (34) in which he paid tribute to the great work Markwick had done as Director. He encouraged new observers to come forward to continue the Section's endeavours stating that he would "be most happy to enlist anyone who is willing to make systematic observations and will gladly give any information and render any assistance in my power". Brook made a similar plea for more observers when he spoke at the BAA meeting a few days later on 29 January 1910 (35). He explained that whilst he did not envisage any major changes to the running of the Section, he hoped to add further stars to its programme over time, something which he did. Like Markwick, Brook firmly believed in encouraging co-operation between observers (3).

Where Brook really excelled was in preparing numerous and timely reports summarising the work of the Section which were published in the Journal. These included three series of annual reports on *SS Cyg*, on "*Long Period Variables*" and on "*Three Irregular Variables*", namely R CrB, U Gem, U Sct (see Table 1 for a list of papers on these three groups of variables). The SS Cyg reports, which will be discussed in more detail later, covered the years 1910 to 1927, and thus continued beyond his time as Director. The other two series covered the years 1910 to 1920. Many of the reports were accompanied by light curves neatly drafted by Brook (he had undertaken similar work for Markwick under his Directorship). During his twelve years as Director, some 83,796 VSS observations were made and discussed, covering approximately 54 stars (33). As examples of his work, light curves of the Long Period Variable X Aur and of R CrB in are shown in Figures 13 and 14.

Brook also published two BAA *Memoirs* on the work of the VSS. The first covers the years 1910-1914 and includes 16,217 observations (36). The second covers 1915-1919 and contains 27,820 observations (37). As Howard Kelly pointed out in his note on the first 50 years of the VSS: "Brook was a stickler for accuracy and he had a keen eye for spotting errors, and the *Memoirs* published during his period as Director are a model of careful and accurate editing. Markwick had previously standardised the presentation of the *Memoirs*, and Brook perfected that which Markwick had initiated" (33). Not only did Brook undertake the hard work of collating the data, analysing the observations and preparing the manuscript for publications, but he also made significant contributions towards the cost of publication from his own pocket. Thus he gave the sums of £50 and £100 pounds towards publication of the two *Memoirs* (approximately 25% of the total cost in each case). Each was a significant amount of money at the time and would be equivalent to several thousands of pounds today. The second *Memoir* was sold to BAA members for 5 shillings (25 p), thus Brook could have bought 400 copies with his donation! He also contributed financially toward the printing of other VSS *Memoirs* published by both his predecessor, Markwick (38), and his successor, Félix de Roy (39). Marwick acknowledged that Brook had also made a significant contribution to compiling the lists of observations published in the earlier *Memoir* during his Directorship. Clearly the BAA, and variable star research in general, has benefitted from Brook's generosity in making these publications available.

Brook retired from the VSS Directorship at the end of 1921. He was clearly highly appreciated and respected by members of the VSS. Twenty-four leading members wrote to the Journal in January 1922 expressing their regret at his departure and putting on "record their high appreciation of his unvarying kindness to each of them, of the great amount of time and trouble which he has bestowed upon their observations, and of the ability with which he has managed the affairs of the Section during the past 12 years" (40). His successor was Félix de Roy (41), a Belgian national living in Antwerp and de Roy immediately launched a collection of contributions from members to purchase a gift with which they could express their appreciation, with a suggested contribution of 10 shillings (50 pence) (42). The gift they settled on was an album "which contains 41 beautiful astronomical photographs" (43). Even after his retirement, Brook remained active in the Section and produced many variable star charts (as well as continuing the series of reports on SS Cyg mentioned previously).

Annie Maunder (1868-1947) (44), writing Brook's obituary some 17 years after his retirement as Director noted (3):

> "Those who came into contact with Mr. Brook agree in stressing two of his characteristics. First, he would approach any new problem slowly, and then after due consideration come to a decision about it that was invariably sound and showed a long view. Second, in spite of being an extremely busy man his correspondence with his observers was voluminous, and no trouble was spared in discussing points of detail that might be of help to the observer or to the work".

**Observations of SS Cygni**

One star in particular which fascinated Brook was SS Cyg and it was the star which he observed most according to the BAA VSS database (28). SS Cyg was discovered in 1896 by Louisa D. Wells of Harvard College Observatory (45) and has been intensively monitored by astronomers around the world ever since. We now know that SS Cyg is a *dwarf nova*. Dwarf novae are semi-detached binaries in which a white dwarf primary accretes material from a secondary star via Roche lobe overflow. The secondary is usually a late-type main-sequence star. Material from the secondary passes through an accretion disc before settling on the surface of the white dwarf. As material builds up in the disc, a thermal instability is triggered that drives the disc into a hotter, brighter state causing an outburst in which the star brightens by several magnitudes (46). In the case of SS Cyg the star spends about 75% of its time in a quiescent state at around magnitude 12.2, but brightens to around magnitude 8.3 every 4 to 10 weeks. Each outburst lasts 1 to 2 weeks (47).

SS Cyg was added to the BAA VSS programme By E.E. Markwick in 1904 (48). Brook started to observe it in 1906 and from 1910 to 1927 he published an annual summary of VSS observations in the JBAA, which was a continuation of Markwick's reports on the star (Table 1). Example light curves, from 1926 and 1927, are shown in Figures 15 and 16. When de Roy became Director in 1922, he asked Brook to continue to analyse and report the observations (49). One aim of the work was to encourage observers to monitor the star as frequently as possible so that all outbursts would be recorded and with such an observational database it was hoped that sufficient data would be available to predict future outbursts. In 1911 Brook wrote:

> "SS Cygni is a fascinating and mysterious star. We cannot hope to explain its changes without a complete record of its history. Towards this end the VSS may claim to have done its share during the past 5 years" (50).

Brook was keen that variable star observers around the world should pool their observations to allow a more complete analysis of the star: "owing to the peculiar character of the variation, it will be difficult for anyone to deal comprehensively with the problem of the star's changes unless the fullest evidence is available" (51). He wrote to Leon Campbell (1881-1951), Recorder of the American Association of Variable Star Observers (AAVSO), with an offer for "the ASVSO [*sic*] and the VSS to compare notes and see if an agreed list of the maxima of this star since 1896 cannot be arrived at" (52). Campbell replied in positive terms, sharing some of the AAVSO conclusions (53).

In spite of the lofty aims of the monitoring programme, by 1914 the true difficulty of the task of predicting the star's behaviour was becoming apparent as Brook commented: (54)

> "It is to be feared that the available material is still insufficient and that we may have to wait some time to predict the star's movements"

Evidently he could not bring himself to believe that the star's behaviour was completely erratic: "the Director thoroughly agrees with Dr. Whittaker [Edmund Taylor Whitaker FRS (1873-1956)] that this cause cannot act wholly capriciously, and if he can solve the mystery it will be a matter of intense interest to those who observe the star" (54). Other people also analysed the available data searching for periodicity. David Gibb, for example, in 1914 found an outburst period of 40.86 days although he admitted that it would probably not be possible to develop an accurate ephemeris, no matter how many terms were included (55). Taking heed of Gibb's comments, Brook conceded in 1915 that: (56)

> "[w]e do not therefore seem to have made much progress, but.....the Director has not the slightest intention of throwing cold water on attempts to penetrate the mystery of the movements of SS Cyg"

Even by 1920 Brook had not given up hope:

> "no evidence has yet been made in predicting its variations; occasional attempts have been made, only to fail. The Section, however, has not lost its interest in the star...and some day we may hope for more light" (57)

By 1926, thoughts of predicting SS Cyg's behaviour had all but faded as Brook commented:

> "I am aware that some attempts have been made to predict, but, as far as I am aware, none have succeeded. Two papers also appeared in [Monthly Notices of the RAS] dealing with the Harmonic Analysis of the star, but neither seemed in any way useful for predicting the future" (58).

We now know that outbursts of SS Cyg are, like all dwarf novae, only *quasi*-periodic, which means they cannot be predicted with certainty. Nevertheless, the observations submitted by BAA VSS members, as well as to other variable star organisations around the world, means that SS Cyg is one on the most intensively monitored of variable stars. As a result, it appears that no outbursts have been missed since it was discovered in 1896 and Brook can be credited for playing his part in encouraging such an intensive study. The accumulated body of data has allowed other researches to conduct studies into the behaviour of SS Cyg. For example, by examining the long term light curve, Leon Campbell was able to categorised outbursts into 4 different classes, depending on the outburst profile (59). A later analysis by Cannizzo and Mattei (60) in 1992 divided the outbursts into type L (long outbursts, exceeding 12 days) and S (short outbursts, less than 12 days) and showed that the most common sequence is LS (134 occurrences), LLS (69 occurrences), LSSS (14 occurrences) and LLSS (8 occurrences). Together these series account for 89% of the outbursts in the study. Cannizzo went on to suggest that the factor which determines the outburst type (L or S) is the amount of material in the accretion disc when the thermal instability is reached (61). Even today, members of the VSS continue to monitor the "ups and downs" of SS Cyg.

**Observations of novae**

The most exciting astronomical event of 1918, and possibly of Brook's Directorship, was the appearance of Nova Aquilae, the brightest nova of the 20$^{th}$ century. There were numerous independent discoveries of the nova, which is now known as V603 Aql, around the world. The first BAA member to have seen it was probably Miss Grace Cook, observing from Stowmarket, on 8 June 1918 at 21.30 UT (62). W.F. Denning (1848-1931) saw it from Bristol at 22.00 UT and Brook himself saw it at Meltham at 22.15 UT. Brook noted

> "[i]t cannot be said to have been "discovered"; anyone who had even the slightest acquaintance with the brighter stars in Aquila could not fail to realise what it was; so the Director, having reached his observatory at [22.15 UT], turned round to glance over the Southern sky and the Nova stared him in the face" (63).

Other BAA members who reported the appearance of the nova that first evening included Mr Packer (22.00 UT) (63), W.H. Steavenson (22.30 UT) (63) (64), Félix de Roy (22.45 UT) (63) (65) and Harold Thomson (23.44 UT) (66) (67). However, there were many claims by people around the world that they had seen the nova several days earlier, which by comparison with observations from reliable observers and from data by photographic sky patrols, were shown to be false (68) (69). In his role as VSS Director, Brook also investigated at least one spurious claim, that of a Captain Piper. Brook's report states that Piper observed "a bright strange star at 12h 45min on 7 June (70), but did not report it because he thought astronomers knew all about it" (71). However, Brook's investigations revealed that a photographic patrol plate taken about 5 hours after Piper's observation showed the nova to be only of the sixth magnitude. On this basis Brook rejected Piper's discovery claim, but rather generously noted that it "must be attributed to a *bonâ fide* mistake in the date" (71).

Brook prepared and distributed a chart to allow observers to estimate the brightness of the nova using a common comparison star sequence, thereby improving the consistency of their observations. Between discovery and 23 July 1918 BAA members reported 392 estimates, of which Brook himself made 18 and Ruth Mary Brook made 13. Brook published a preliminary report in the September 1918 Journal, along with a light curve (71). He remarked on the superficial similarity of the outburst behaviour of the nova to the dwarf nova SS Cyg, but pointed out that nova increased in brightness far more rapidly (71) (72).

Several observers reported rapid changes in brightness as they watched the star, including Grace Cook, E.E. Markwick, Walter Goodacre (1856-1938) and 3 other observers. By contrast, according to Brook, "Mr Steavenson and others do not believe in the reality of these short period changes" (71). Félix de Roy also appeared to doubt the validity of his observation:

> "I don't believe in *real* short period fluctuations of the light of the Nova, but I think the aspect was peculiar. What this peculiarity *is* seems difficult to describe; it was something more glowing and piercing that an ordinary star, and quite independent of colour".

Brook himself reported (71) that the visual aspect of the nova was strange, saying that "it seemed larger than its brightness warranted". He noted that R.L. Waterfield (1900-1986) thought that variations of 0.2 to 0.3 magnitude were possible, but Brook went on to say that "[t]his would entail almost momentary variations of 25 to 30 per cent of the total integrated light, not of a "fire flickering", but of an enormous volume of glowing gas, hundreds of thousands of miles in diameter. *It is difficult to believe this is possible*" (the italics are mine) (71). Was it Brook's sheer incredulity that such a phenomenon could exist that prevented him from reporting it? Given that the phenomenon was reported by at least 6 observers in Nova Aquilae, it seems likely that it was real.

Brook also observed and reported on at least two other novae during his Directorship. He gave a talk on the observations that he and other members of the VSS had made of Nova Geminorum 1912 (later named DN Gem) at the April 1912 BAA meeting (73). The nova reached third magnitude and initially it was yellowish white, becoming orange as it faded. Unusually, his observations of the nova from Meltham were facilitated by the ongoing national coal strike (74): "we are practising severe economy in coal consumption in the north, and consequently every gas lamp in the valley has been extinguished, a result clear to the heart of a star-gazer" (75) (in those days town gas was derived from coal). Given the industrial nature of Meltham at that time, with the large mill complex, one wonders to what extent local air pollution may have affected Brook's observational work. In December 1904, while attempting to observe an occultation of a Jovian satellite, he noted that "[a] good deal of smoke-fog and poor definition made exact time of observations difficult" (76).

In 1920 W.F. Denning discovered a nova in Cygnus (Nova Cyg III 1920 or V476 Cyg) (77), a large amplitude and extremely fast nova, increasing in brightness by nearly 13.5 magnitudes and then declining 3 magnitudes in 16 days. Brook's analysis and light curve (Figure 17) of the early stages of the outburst were presented in the *Journal* (78).

The BAA meeting of 27 March 1901 was largely given over to papers and a discussion on Nova Persei 1901 (GK Per). The nova had been discovered on the evening of 21 February 1901 by the clergyman Thomas David Anderson (1853-1932) when he was walking home in Edinburgh. It reached its brightest two days later at magnitude 0.2. Six days after maximum, the nova had faded to second magnitude and two weeks later it reached the fourth. During the BAA meeting Brook shared his observations of the nova including some basic spectroscopic studies made with a McClean spectroscope in combination with his 9 inch reflector.

Brook also became involved in discussing a mystery objected observed by E.E. Barnard (1857-1923) in 1892 (79). Barnard was observing Venus with the 36 inch Lick refractor in the morning twilight on August 13, when he noticed a star of about the seventh magnitude in the same field. Although he was unable to identify the star after consulting a star atlas, he did not report the observation until 1906 (80). On reading an abstract of this paper in The Observatory, Brook speculated "[T]he position of Venus on August 13, 1892, was, according to *Whitaker's Almanac*, R.A. 6h 52 m 40 s, Dec. +17º 12´. This places it close to (perhaps within) the border of the Milky Way in Gemini; we know that new stars, almost without exception, have appeared in or near the Milky Way; is it not possible that what Prof. Barnard saw was a Nova?". The mystery has never been resolved. Barnard is renowned as a careful observer, so there is no reason to doubt his observation. It is entirely possible that Barnard indeed saw a short-lived nova as Brook had suggested. The object is listed in the New Catalogue of Suspected Variables as NSV 3313.

**The total solar eclipse of May 1900**

The track of the 28 May 1900 solar eclipse passed through the southern USA, across the Atlantic, reaching land in Portugal, then onwards through Spain, over the Mediterranean and into Algeria. Members of the BAA observed the eclipse from each of these countries (81) (as well as at sea off Portugal) and a report of their expeditions was published in a BAA *Memoir* (82), a page of which is shown in Figure 18, edited by E.W. Maunder (1851-1928; Figure 19). Moreover, upon their return to Great Britain, several members of the Association, including Brook, presented reports at the BAA meeting on 27 June 1900 (83) (84).

Brook travelled to Algeria with his sister Ruth Mary Brook (Figure 3). Ruth, usually referred to as "Mrs Arthur Brook" in the *Memoir*, was the widow of Arthur Brook (1857-1888) (85). Ruth herself also had an interest in astronomy and joined the BAA in 1900 (86). She served on BAA Council 4 times between 1913 and 1918 (87). Like her brother, she observed variable stars and her observations of $\chi$ Cyg, W Cyg and R Leo were reported in VSS *Memoirs* (36) (37). The Brooks met up with E.W. Maunder's eclipse party at Marseilles where they boarded the steamer *General Chanzy*. The party included Maunder's wife Annie (Figure 20), who later wrote C.L. Brook's RAS obituary, their two daughters (88), and Andrew Claude de la Chalois Crommelin (1865-1939) and his wife (89). Other members of the BAA also observed the eclipse in Algiers, including R.F. Roberts, who was killed in an air raid on London during World War One (90). The party's base in Algiers was the Hôtel de la Régence which afforded a flat roof from which they observed the eclipse in its full glory (Figure 21) (91). E.W. Maunder noted that "[o]ur hotel was in the very centre of the city, facing its chief *Place*, a site which in a northern clime would not be ideal for an observing station, but which here in smokeless, fireless, subtropical Algiers had few drawbacks and not a few advantages........its roof was thoroughly well adapted for our requirements in an observing station" (82). The various chimney stacks on the rook were used as piers to support telescopes (Figure 22), although "their most serious defect was in presence of the vent, down which it was so easy to drop eyepieces and screws and other useful or indispensable articles" (91). In the days before the eclipse, there were many excursions from the hotel to visit the observing stations of other eclipse viewers in and around Algiers (92). Similarly, many visitors also came to inspect the roof-top observation area of the Hôtel de la Régence. One such visitor was Princess Amelie of Schleswig-Holstein, aunt of the Empress of Germany, who was also staying in the Hôtel de la Régence (93). A further visitor was the colourful adventurer Leo Brenner of the Manora Observatory on the Istrian coast (94).

The entirety of the eclipse passed off under clear skies; Crommelin was charged with observing the early stages and reporting progress to the rest of the party, informing them totality was about to commence. He noted that "Mr Brook kindly gave assistance in noting the time of first contact" (82). Everyone was able to see the corona during totality and several members obtained photographs. A drawing of the eclipse is shown in Figure 23.

The only optical aid Brook and his sister used during the eclipse was apparently binoculars. Brook himself noted the time of appearance of stars as the sky faded towards totality (95). During totality he could see Castor and Pollux, Aldebaran and Mercury with the naked eye; "[m]y impression is that during totality I could have seen all second magnitude stars, or even 2½ magnitude stars, provided that they had been some distance above the horizon.....Many stars escaped being seen because there is no time to search for them" (82). Totality lasted about a minute.

The Brooks also made meteorological measurements and observations of shadow bands during the eclipse and Brook himself contributed a detailed report which appeared in the BAA eclipse *Memoir*. To facilitate detection of the shadow bands, they spread a large white sheet on the hotel roof upon which were sewn two concentric circles of black tape (the sheet can be seen in Figure 21). They hoped to use the circles to count the number of bands in a given distance, but this turned out not to be possible as the bands were rather indistinct, in reality being low contrast ripples (83). As Brook himself noted:

> "the word 'bands' was quite inapplicable to what they saw; the shadows were more like "ripples" on water" (83)

The "bands" appeared about 3¼ minutes before totality and moved at 1½ yards/second (~1.4 metres/second). Immediately at the end of totality, Ruth Brook reported that the appearance of shadows on the sheet as "being covered with dark blotches of shadow which appeared to be in a state of violent agitation" (83). The appearance of these shadow patches, which lasted 7 or 8 seconds, is shown in Figure 24. They were noted immediately after the end of totality, but were apparently not visible leading up to it.

Brook set up a meteorological station comprising various wet and dry bulb thermometers, a Stevenson screen and a wind vane. Brook's temperature measurements made during the eclipse are shown in Figure 25, although he commented "I am not quite sure of the value of the thermometer observations during an eclipse; they do not seem to have much bearing on eclipse phenomena" (82).

**Other astronomical observations**

Although Brook mainly concentrated on observing variable stars, other subjects did not escape his scrutiny completely, especially in the years before he became VSS Director. These included Jupiter's satellites (76) (96), eclipses, meteors and the zodiacal light. He observed a partial eclipse of the Sun from Meltham on 28 June 1908 during which "there was a persistent impression that a *very small* portion of the edge of the Moon could be seen outside the Sun's disc at both cusps" (97).

In November 1904 he observed the Leonid meteor shower, finding the display was "very feeble" on the night of 14 to 15 November, whereas the following night the display was "superior" (76). He observed a fireball on the evening of 16 November, which he described "as much brighter than 1st mag. It gave a green flash and left a short streak amongst the stars low in the S.S.E. for 1 minute.....duration of flight 1 second", according to a report in *The Observatory* compiled by W.F. Denning (98). The report states that the fireball was also seen by three people at Greenwich and "its apparent brilliancy much exceeded that of Jupiter. Near the termination of its flight it burst and lit up the entire sky". Remarkably, his sister, Ruth Mary Brook, observed the same fireball from Charmouth in Dorset. The three separate observations allowed Denning to compute the trajectory of "not more than 60 miles" (97 km) from a height of 88 to 44 miles (142 to 71 km) from near Petersfield (Hampshire) to Hungerford (Berkshire).

Brook regularly recorded the presence of the zodiacal light and the gegenschein. For example, in 1901 he saw the gegenschein on at least 8 nights from Meltham and several other times in 1902 (99). Subsequently, he saw the zodiacal light three times whilst staying on the Isle Wight in March 1903 which he described at the BAA meeting in London the following week (100). We must conclude from the Meltham observations that the skies were often very clear and dark there, in spite of the problems with local air quality experienced at other times which were referred to previously.

On the evening of 9 February 1907 Brook was impressed by a spectacular aurora, which pulsated with reds and greens:

> "This is the most remarkable auroral display I have ever seen; it may not have been as bright as that of 17$^{th}$ November 1882,.....but I have never witnessed anything even approaching the extraordinary activity of the light pulsations as seen between 10.10 and 11.00 pm" (101)

The next day he looked at the sun with his naked eye, using a dark glass, and "saw very distinctly three large sunspots......On using a binocular many other smaller spots were seen".

**Meteorology**

Apart from astronomy, Brook's other great passion was meteorology (4), as exemplified by his meteorological measurements during the 1900 solar eclipse. He was elected as a Life Fellow of the Royal Meteorological Society on 18 February 1891 and he served on its Council in 1913 and as Vice President in 1922 (just after he resigned as VSS Director). He was a contributor of measurements to the British Rainfall Organisation (102), which he also aided financially, and he served on the Organisation's board of trustees from 1909 until 1919 when control of the Organisation passed to the Meteorological Office. Some years later, when the Thunderstorm Census Organisation (103) was formed, Brook was again a generous financial supporter (4).

Brook began regular climatological records at Meltham in 1878 and maintained them until his death. He had a particular interest in the link between atmospheric phenomena and astronomy, writing on such subjects as lunar halos (104), lunar rainbows (105) (106) and the presence of volcanic dust in the atmosphere leading to dark lunar eclipses (107), as well as more mainstream meteorological topics such as severe snowfalls (108).

**Conclusion**

The BAA enjoys an international reputation for the quality of the observational work carried out by its members for more than 120 years. Charles Lewis Brook played an important part in developing that reputation in the work he did whilst Director of the VSS. He was not merely interested in collecting the observations of the members (to which he also contributed), but he spent considerable amounts of time analysing the data and writing numerous papers on the findings. These publications, which in many cases he generously funded from his own pocket, made the results of many amateur astronomers available to the astronomical community and the data they contain are of lasting value. He deserves to be long remembered in the history of the BAA for his contributions to the Association.

**References and notes**

1. Meltham is located within the Metropolitan Borough of Kirklees, West Yorkshire. It lies in the Holme Valley, below Wessenden Moor, four and a half miles south-west of Huddersfield on the edge of the Peak District National Park.

2. Lindley W.M., JBAA, 50, 33 (1939).

3. Maunder A.S.D., MNRAS, 100, 233 (1940).

4. Obituary (author not specified), Quarterly Journal of the Royal Meteorological Society, 65, 471 (1939).

5. Unknown author. I have not been able to obtain a copy of the original published obituary from the *Huddersfield Examiner* of May 1939. However local historian, Mr. Duncan Crawshaw of St. Bartholomew's Church, Meltham, provided me with a type-written copy.

6. According to local historian, Mr Duncan Crawshaw, as a child Brook lived in Thickhollins Hall, Meltham, which for many years was the family seat. Thickhollins is now occupied by Meltham Golf Club.

7. Rev. Lewis Jones was involved in some controversy in 1828. In that year the Vicar of Meltham, Rev. Edmund Armistead, died and the people of Meltham favoured his curate, Robert Kean, as successor. In spite of their wishes, the Vicar of the neighbouring Parish of Almondbury, Lewis Jones, was appointed instead. When Jones arrived in Meltham, Kean barricaded himself into the Church, refusing to give up his place in the pulpit even when Jones broke into the Church with the help of a contingent of special constables. Riots ensued and some men were arrested, later being bound over to keep the peace. Eventually the church courts found in favour of Jones who became vicar for the next 10 years. These events described in the booklet "Worship for all: a history of public worship in Meltham", available from St. Bartholomew's Church, Meltham.

8. Crawshaw D., Personal communication (2010).

9. Brook boarded at The Lodge Boarding House at Uppingham. The Headmaster at the time, serving from 1853-1887, was Edward Thring (1821-1887) who developed an international reputation for his pioneering educational ideas, including installing the first gymnasium in an English school and he later added a heated indoor swimming pool.

10. Tyers N., Old Uppingham Association, personal communication, from the archives of Uppingham School (2010).

11. Brook was also in the Trinity College 1$^{st}$ eleven. The obituary in the Huddersfield Examiner notes that "Mr. Brook...was known as a fielder of outstanding ability. He could throw a ball perfectly with either hand".

12. Harewood Lodge was pulled down in the 1980s and is now the site of a Church near the junction of Huddersfield Road and Meltham Mills Road (The Church of Jesus Christ of Latter Day Saints). I have spoken to several Meltham residents who were associated with the Lodge after World War Two, some even living there after it was converted to flats, but none can recall the site of an observatory there. .

13. The foundation stone and memorial to Brook's father can be found inside Helme Church (rather ungrandly, it is located inside the church WC as a result of more recent internal changes in the Church building). "This Church was built and endowed in the memory of Charles John Brook of Thickhollins by his brothers and sisters. ....This stone was laid.....by his only son". His only son being, of course, Charles Lewis Brook.

14. Records of Coats Viyella PLC, thread manufacturers, Paisley, held at Glasgow University Archive Services, reference GB 0248 UGD 199/1, 28, 32.

15. Records of Jonas Brook & Bros. Ltd., thread manufacturers, Meltham Mills, Huddersfield, held at Glasgow University Archive Services, reference GB 0248 UGD 199/18.

16. Records of James Chadwick & Bros. Ltd., cotton manufacturers, Bolton, held at Glasgow University Archive Services, reference GB 0248 UGD 199/26.

17. The goat's head emblem of Jonas Brook & Bros. Ltd is the emblem of the Meltham & Meltham Mills Brass Band, which was established in 1846. The band enjoyed support of the various Brook businesses and is still active as a brass band today. Being the Brook family coat of arms, the emblem is also in St. James' Church at Meltham Mills, with the motto "en dieu ma foy" (in God we trust).

18. Thread manufacture at Meltham Mills went into sharp decline after the First World War. David Brown Tractors took over the factories in 1939 and built tractors there, finally closing operation in 1988. Meltham Mills is now divided into a series of separate units occupied by various light industrial and service companies.

19. Meltham residents still speak of the generous donations Brook made. As an example, on 20 Feb 1936 he donated £400 towards the cost of recommissioning the bells at St. Bartholomew's Church, Meltham as recorded in. "The Parish Church of St. Bartholomew Meltham: tercentenary 1651 to 1951", by J.E. Roberts (1951).

20. For many years, Meltham Mills Cricket Club played in the Huddersfield & District League and in 1895 they headed the league. They enjoyed much financial support from the Brook family. In 1925 a new cricket pavilion was opened, the gift of Jonas Brook & Bros. The club was dissolved in 1940, following the closure of the thread mills. For further details see: http://www.ckcricketheritage.org.uk/MMHistory.pdf.pdf

21. The obituary in the *Huddersfield Examiner* states that the funeral service was conducted by Archdeacon A. Baines and Canon H.F.T. Barter (Vicar of St. Bartholomew's, Meltham). The "family and personal mourners were. Col. C.J. Brook (nephew) and Mrs. Brook, Col. C.J. Hirst (nephew), Mrs. Fisher (niece), Miss D. Brook (niece), Mr. T.W. Hirst (nephew), Lady Darlington (niece), Col. and Mrs. Simpson, Admiral Watson, Miss Watson, Mr. E.W. Brook, Mrs. Crabbe, Maj. T. Brooke, Mr. Harry Armitage, Mrs. Watson and Mr. Lart.

22. Brook's RAS Fellowship application form was countersigned by Thomas K. Mellor FRAS and H.F. Newall. Prof. Newell was for a time Director of the Solar Physics Laboratory at Cambridge, where he specialised in spectroscopy. Newell observed the total solar eclipse at Algiers in 1900, which Brook also observed and which is discussed later in this paper. However, they observed from different locations, with Newall being located in the grounds of the Algiers Observatory; see Turner H. H. and Newall H. F, Proc. Royal Society of London, 67, 346-369 (1900)

23. Chapman A., in "*The Victorian Amateur Astronomer*", John Wiley & Sons (1998).

24. Toone J., JBAA, 120, 135 (2010).

25. JBAA, 9, 347 (1899). "List of people seeking election as members of the Association". The proposer was Thomas K. Mellor (who also supported his RAS Fellowship) and the seconder was J.G. Petrie, who was BAA Secretary at the time (serving 1894-1907).

the Boer War which was in progress. at the time; see Chapter 5 of "The British Astronomical Association, the First Fifty Years"

85. It appears that Arthur Brook, who was originally from Essex, was no relation of the Brooks of Meltham.

86. Mrs Arthur Brook was elected to the Association on 19th Dec 1900 [JBAA, 11, 139 (1901)]. The proposer and seconder were Irene M. Maunder and Catherine A. Stevens respectively [JBAA, 11, 96 (1901)]. Irene Matilda Maunder (1880-1977) was the daughter of E.W. Maunder and his first wife, Edith Hannah Maunder, and a member of the BAA (1899-1907). For further details see: Kinder A.J., JBAA, 118, 21 (2008)

87. Sessions 1913-14, 1914-15, 1915-16, 1916-17; "Minutes Book of Annual, Extraordinary, Special Meetings 1910-1968", British Astronomical Association.

88. Maunder's daughters were Edith Augustus (1878-1966) and Irene Matilda (1880-1977), who signed Ruth Mary Brook's BAA membership application as noted previously.

89. Crommelin, from the Royal Greenwich Observatory, went on to become the President of the BAA 1904-1906. He was Director of the Comet Section 1897-1939. The eclipse was also observed from Algiers by two of Brook's associates, Prof. H.H. Turner and H.F. Newall, although apparently they travelled independently of Maunder and Brook.

90. R.F. Robert's daughter was also killed in the air raid; see Chapter 5 of "The British Astronomical Association, the First Fifty Years". Roberts can be seen in the photograph taken on the hotel roof in Algiers, which appears in this paper.

91. Maunder also pointed out in the Memoir that a further advantage of the roof was that the observers were separated from the throng of people in the streets that might have disturbed their observations. Maunder commented that there were 20,000 or 30,000 people within sight of the roof-top. The hotel waiters also expressed patriotic concern on learning that "the eclipse came to Spain before it passed through Algeria. and that it was total longer there"

92. One such observer was John Evershed (1864-1956) who had set up an encampment some 20 miles from Algiers to observe the flash spectrum. At night the eclipse party at Hôtel de la Régence took advantage of the clear skies to observe southern constellations. including Centaurus, Scorpio and Sagittarius, as well as to adjust their telescopes

93. Princess Amelie sailed to Algiers on the "Argonaut". She invited Maunder and his wife to visit her there to "explain the circumstances of the eclipse".

**Acknowledgements**

I am most grateful for the assistance I have received from many people during my research for this paper. The Rev. Maureen Read, Mr. Duncan Crawshaw and Mrs. Mary Crawshaw kindly hosted me for a day at Meltham during which they showed me around the Meltham Churches and other landmarks associated with Brook, as well as providing interesting pieces of background information on the Brook family. Their local knowledge helped me to understand more fully the impact of the Brook family on life in Meltham. Hugh Darlington, great grandson of Esther Frost Brook (Charles Lewis Brook's sister), provided the photographs of Charles Lewis and Ruth Mary Brook used in the paper – until Hugh contacted me I had been unable to find any photographs of the family, so his contribution is highly valued. Nigel Homer (109) kindly gave me permission to use his photograph of St. James' Church, Meltham Mills. Mike Saladyga (AAVSO) provided copies of letters between Brook and Leon Campbell from the AAVSO archives. Peter Hingley (RAS) looked after me on numerous visits to the RAS library and provided a copy of Brook's RAS Fellowship application form. John Toone advised on various historical matters pertaining to the BAA VSS. Roger Pickard provided details of Brook's BAA VSS observations from the VSS database. Nicola Tyers (Uppingham School, Old Uppingham Association) provided information on Brook's school days. This research made use of SIMBAD, operated through the Centre de Donées Astronomiques (Strasbourg, France), and the NASA/Smithsonian Astrophysics Data System. Finally I would like to thank both referees, Roger Pickard and Mike Frost for their helpful comments that have improved the paper.

**Address**

"Pemberton", School Lane, Bunbury, Tarporley, Cheshire, CW6 9NR, UK [bunburyobservatory@hotmail.com]

| Title | Reference |
|---|---|
| SS Cygni in 1910 | (50) |
| SS Cygni in 1911 | (110) |
| SS Cygni in 1912 | (51) |
| SS Cygni in 1913 | (54) |
| SS Cygni in 1914 | (56) |
| SS Cygni in 1915 | (111) |
| SS Cygni in 1916 | (112) |
| SS Cygni in 1917 | (113) |
| SS Cygni in 1918 | (114) |
| SS Cygni in 1919 | (57) |
| SS Cygni in 1920 | (115) |
| SS Cygni in 1921 | (49) |
| SS Cygni in 1922 | (116) |
| SS Cygni in 1923 | (117) |
| SS Cygni in 1924 | (118) |
| SS Cygni in 1925 | (58) |
| SS Cygni in 1926 | (119) |
| SS Cygni in 1927 | (120) |
| Long Period Variables in 1910 | (121) |
| Long Period Variables in 1911 | (122) |
| Long Period Variables in 1912 | (123) |
| Long Period Variables in 1913 | (124) |
| Long Period Variables in 1914 | (125) |
| Long Period Variables in 1915 | (126) |
| Long Period Variables in 1916 | (127) |
| Long Period Variables in 1917 | (128) |
| Long Period Variables in 1918 | (129) |
| Long Period Variables in 1919 | (130) |
| Long Period Variables in 1920 | (131) |
| Three Irregular Variables in 1910 | (132) |
| Three Irregular Variables in 1911 | (133) |
| Three Irregular Variables in 1912 | (134) |
| Three Irregular Variables in 1913 | (135) |
| Three Irregular Variables in 1914 | (136) |
| Three Irregular Variables in 1915 | (137) |
| Three Irregular Variables in 1916 and 1917 | (138) (139) |
| Three Irregular Variables in 1918 | (140) |
| Three Irregular Variables in 1919 | (141) |
| Three Irregular Variables in 1920 | (142) |

**Table 1: Brook's JBAA papers on SS Cyg, Long Period Variables and Irregular Variables**

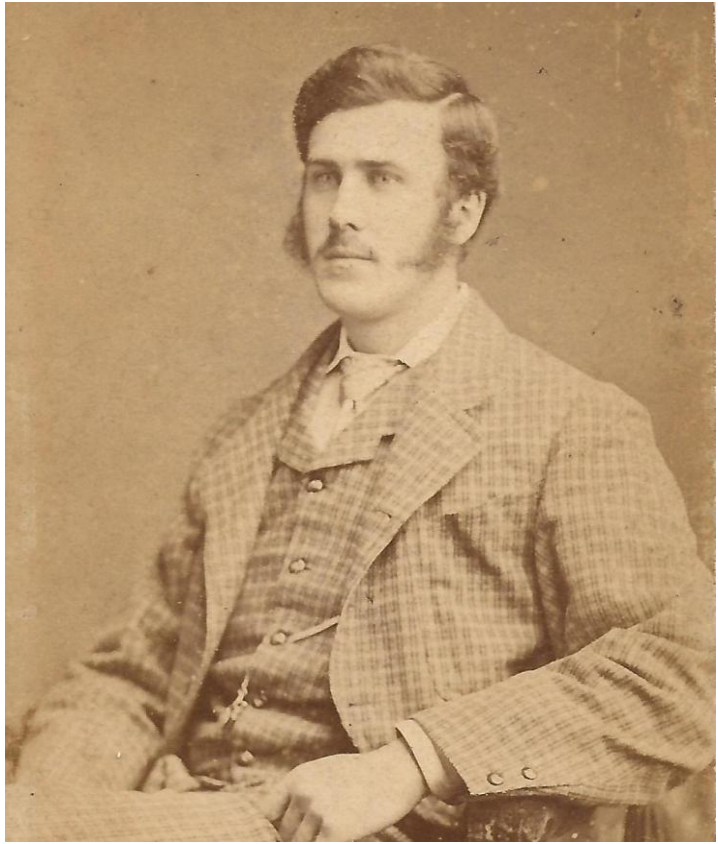

**Figure 1: Charles Lewis Brook**

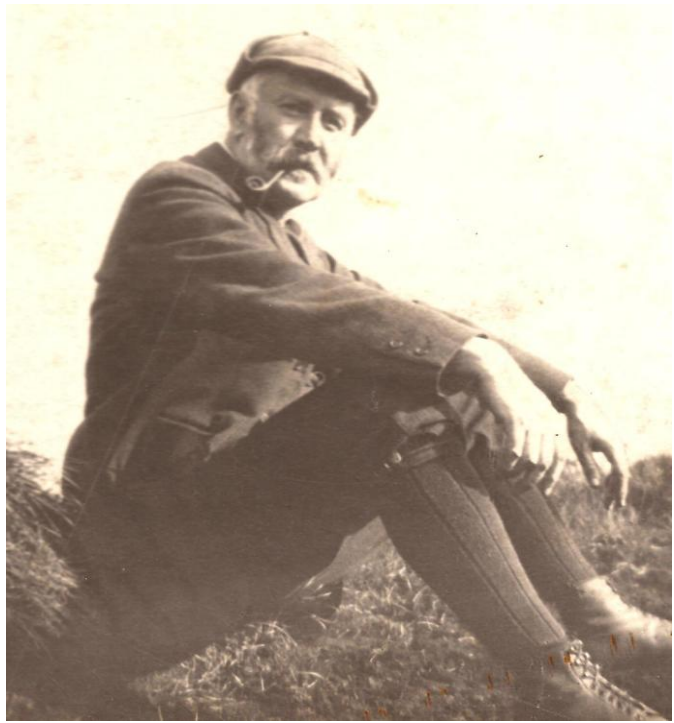

**Figure 2: Charles Lewis Brook in later years**

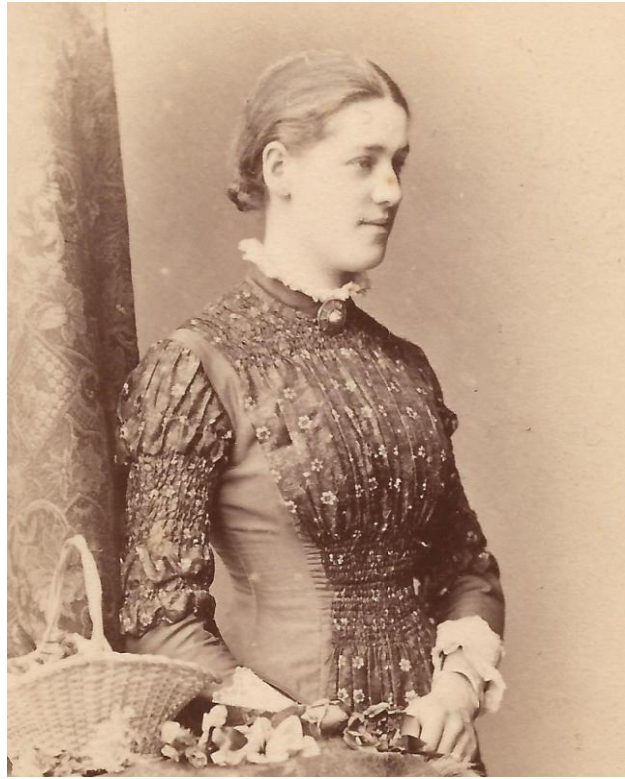

**Figure 3: Ruth Mary Brook**

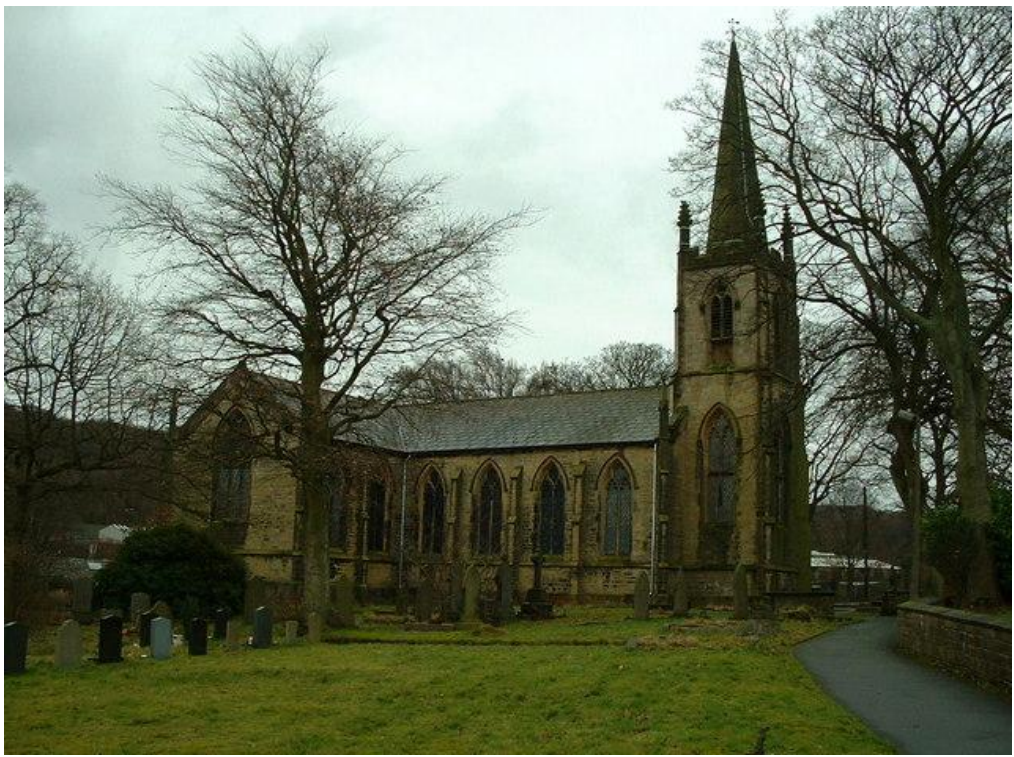

**Figure 4: St. James' Church, Meltham Mills**

(Nigel Homer)

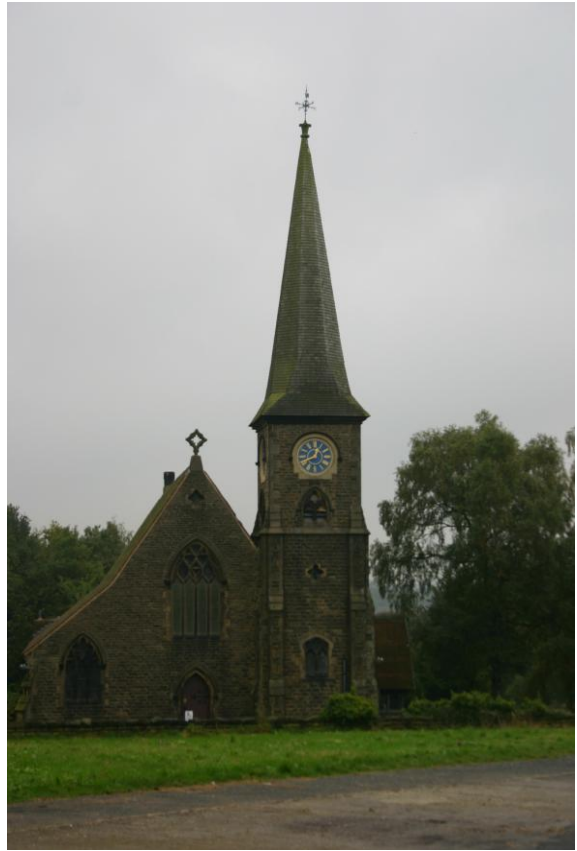

**Figure 5: Christ Church, Helme**

(Jeremy Shears)

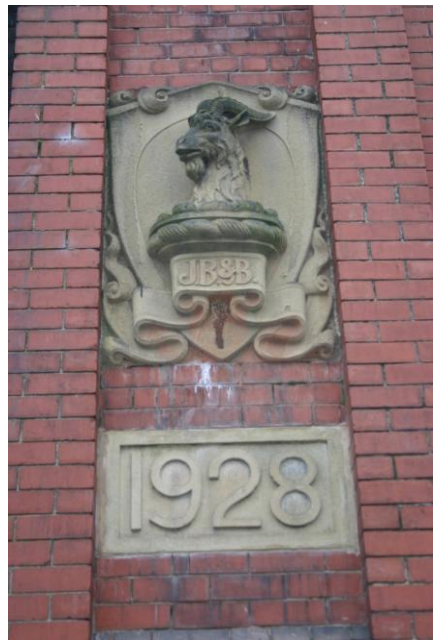

**Figure 6: The goat's head logo of Jonas Brook & Bros**

(Jeremy Shears)
This logo appears on one of the mill buildings in Meltham Mills

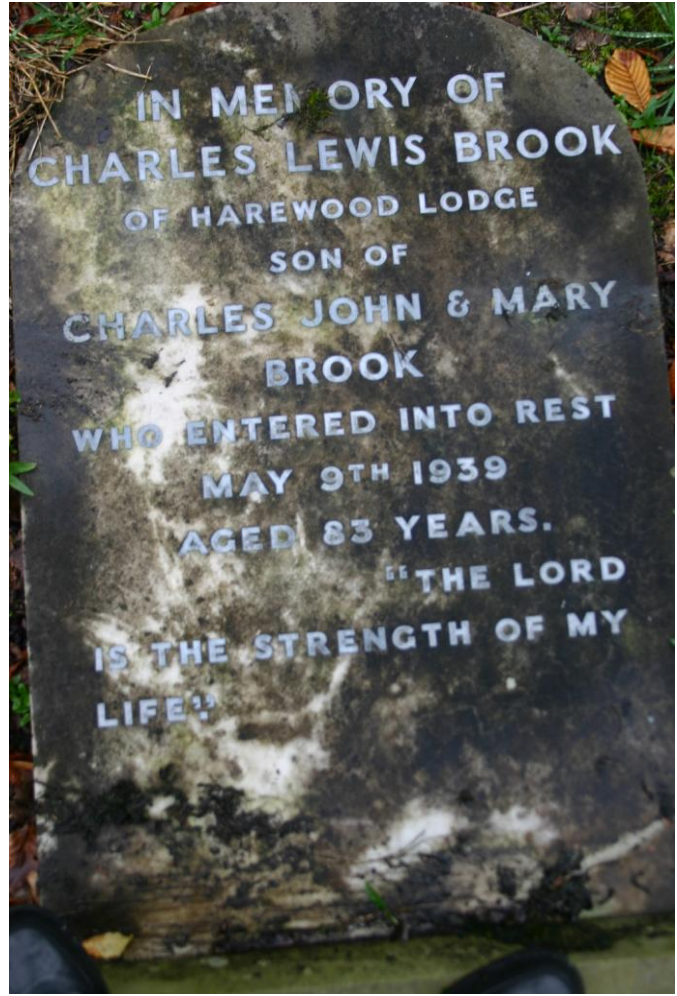

**Figure 7: Charles Lewis Brook's gravestone in the churchyard of St. James' Church, Meltham Mills** (Jeremy Shears)

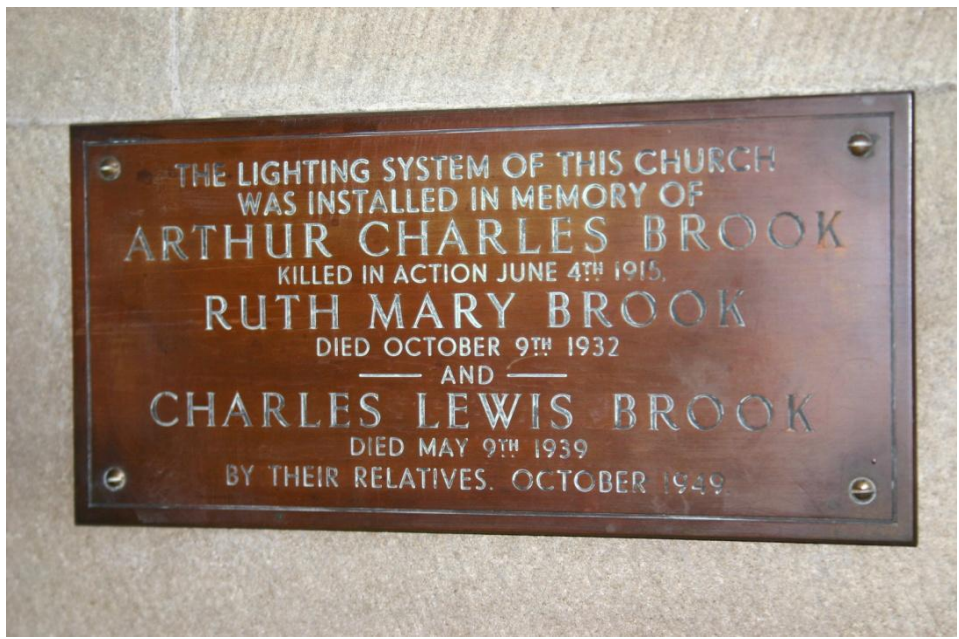

**Figure 8: Brook family memorial in Christ Church, Helme** (Jeremy Shears)

*Accepted for publication in the Journal of the British Astronomical Association*

ROYAL ASTRONOMICAL SOCIETY.

Charles Lewis Brook MA F.R.Met.Soc.

of Harewood Lodge Meltham, Huddersfield

being desirous of admission into the ROYAL ASTRONOMICAL SOCIETY, We, the undersigned, propose and recommend him as a proper person to become a Fellow thereof.

Witness our hands, this 25th day of October 1898

    T. H E C Espin { from personal knowledge.

    Thos. K. Mellor.

    H. F. Newall.

Proposed 1898 Nov. 11.
Elected 1899 Jan 13

**Figure 9: Brook's RAS application form**

ROYAL ASTRONOMICAL SOCIETY.

*Form to be subscribed by every Fellow previously to his Admission.*

I, the undersigned, being elected a Fellow of the Royal Astronomical Society, do hereby promise that I will be governed by the Charter and Bye-laws of the said Society, as they are now formed, or as they may be hereafter altered, amended, or enlarged; that I will advance the objects of the said Society as far as shall be in my power: Provided, however, that whenever I shall signify in writing to the Society that I am desirous of withdrawing my name therefrom, I shall (after the payment of any annual contribution which may be due by me at that period, and after giving up any books, instruments, or other property belonging to the Society, in my possession, or intrusted to me), be free from this obligation.

Witness my hand, this 16th day of January 1899.

Charles Lewis Brook. M.A. F.R.Met.Soc

Harewood Lodge
Meltham
" Huddersfield
Yorkshire.

**Figure 10: Brook's RAS Fellowship form**

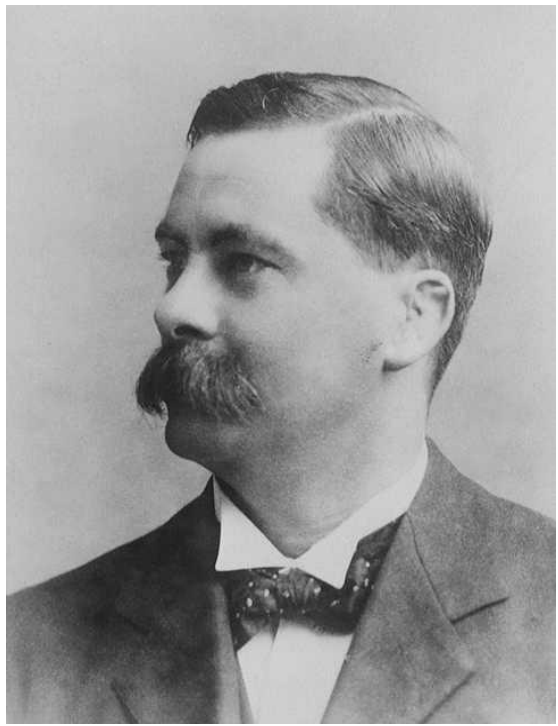

**Figure 11: Prof H.H. Turner (1861-1930)**

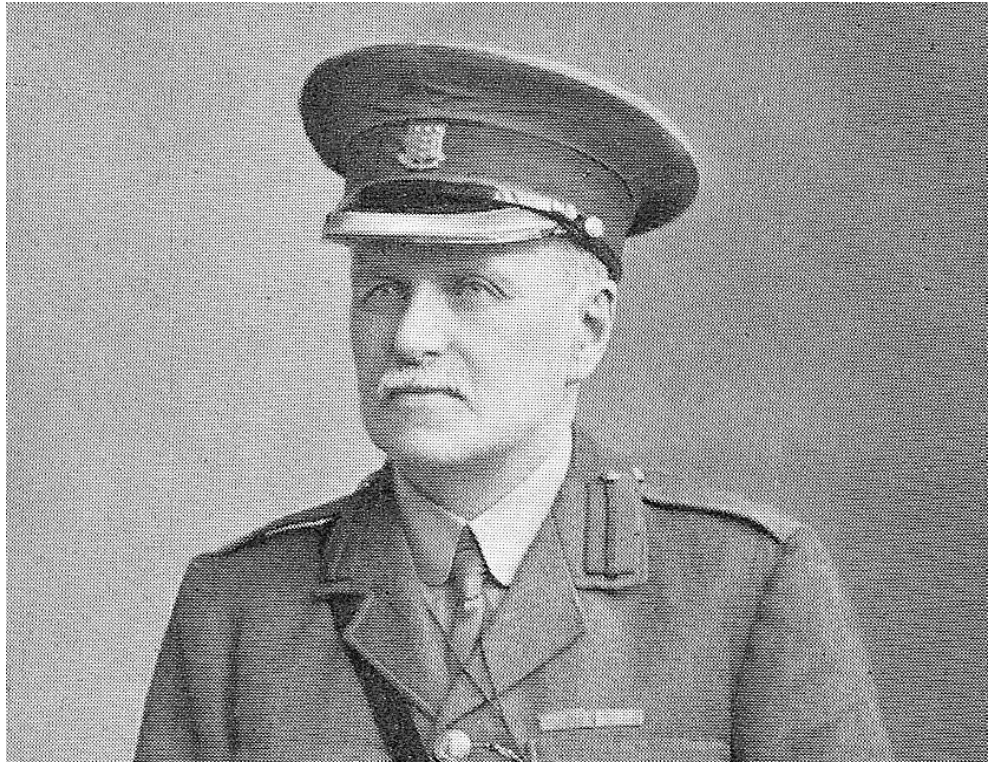

**Figure 12: E.E. Markwick (1853-1925)**

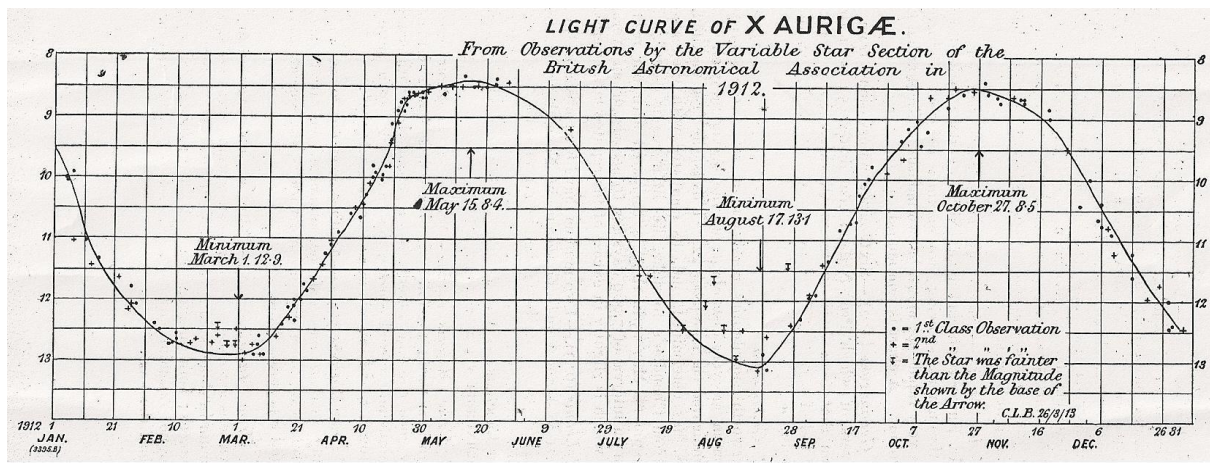

**Figure 13: X Aur in 1912**

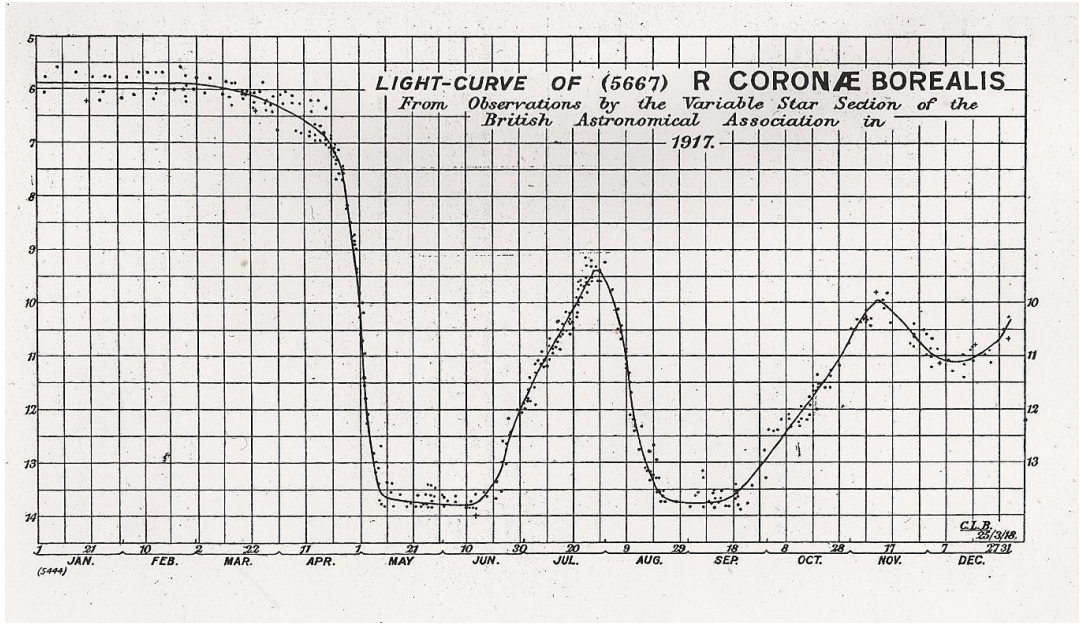

**Figure 14: R CrB in 1917**

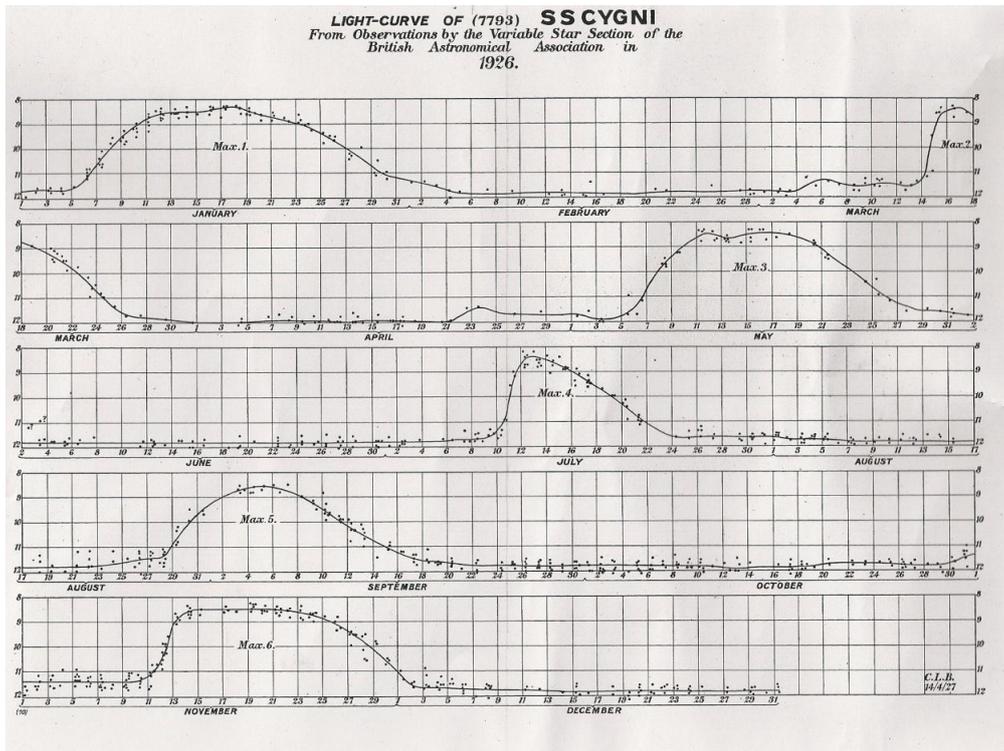

**Figure 15: SS Cyg in 1926**

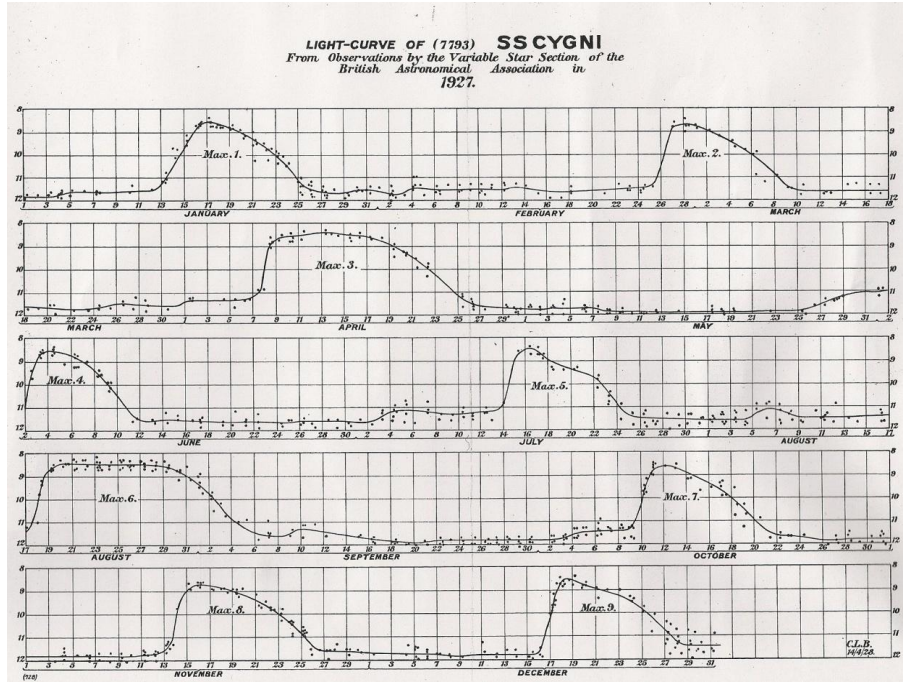

**Figure 16: SS Cyg in 1927**

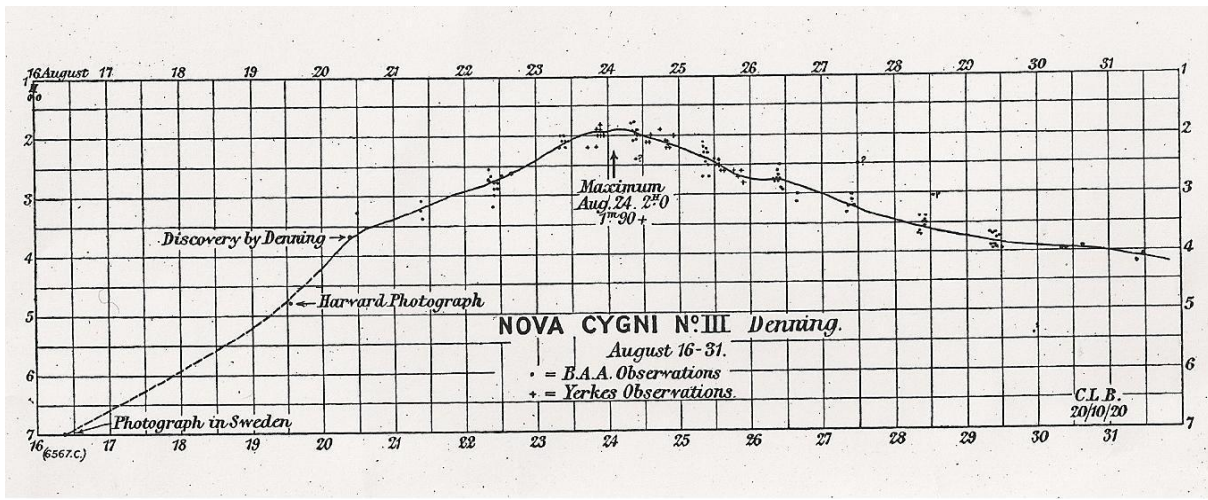

**Figure 17: Nova Cyg III in 1920**

Figure 18: The BAA Memoir of the 1900 solar eclipse (82)

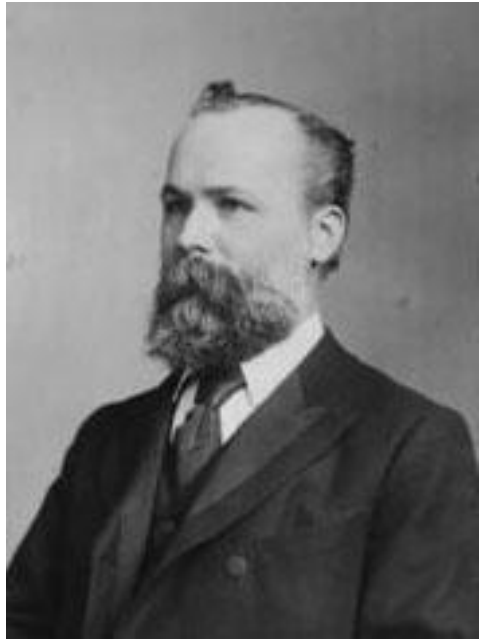

**Figure 19: E.E. Maunder (1851-1928) photographed in 1905**

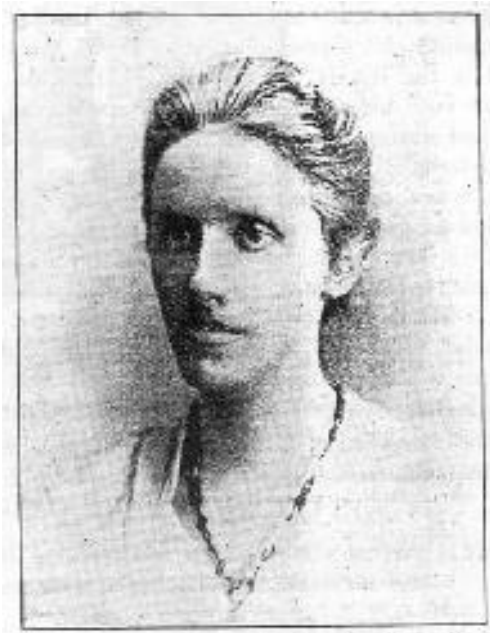

**Figure 20: Annie (A.S.D.) Maunder (1868-1947)**

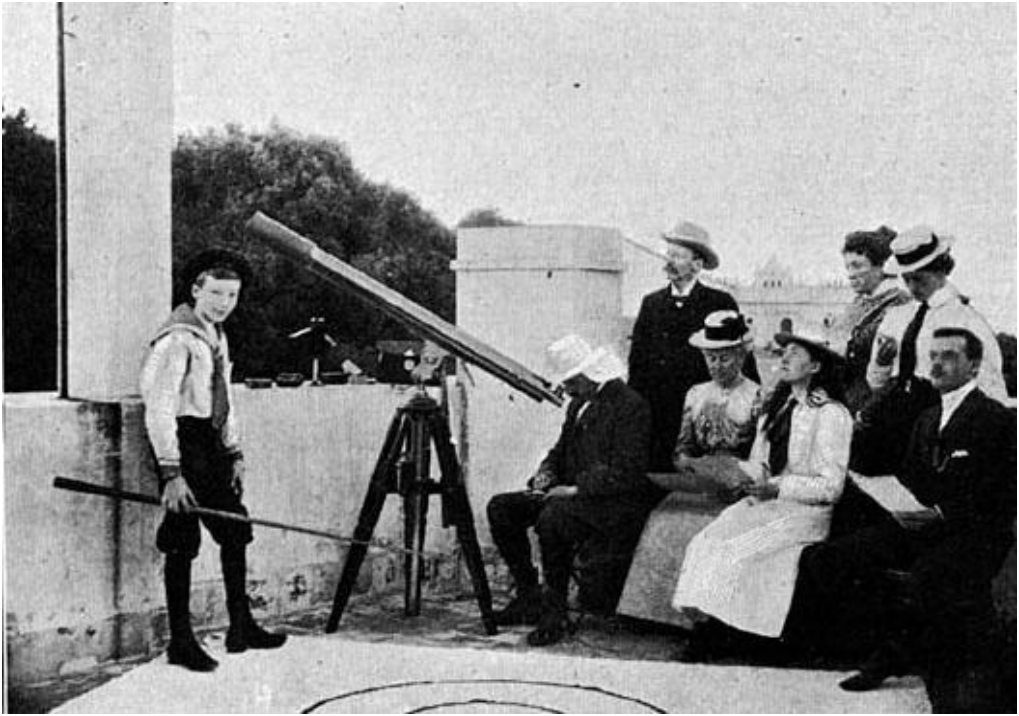

**Figure 21: Eclipse observers on the roof of the Hotel de la Régence, Algiers** (82)
(143)

Note the sheet with two concentric circles brought on the expedition by the Brooks for monitoring shadow bands and which is discussed in the text

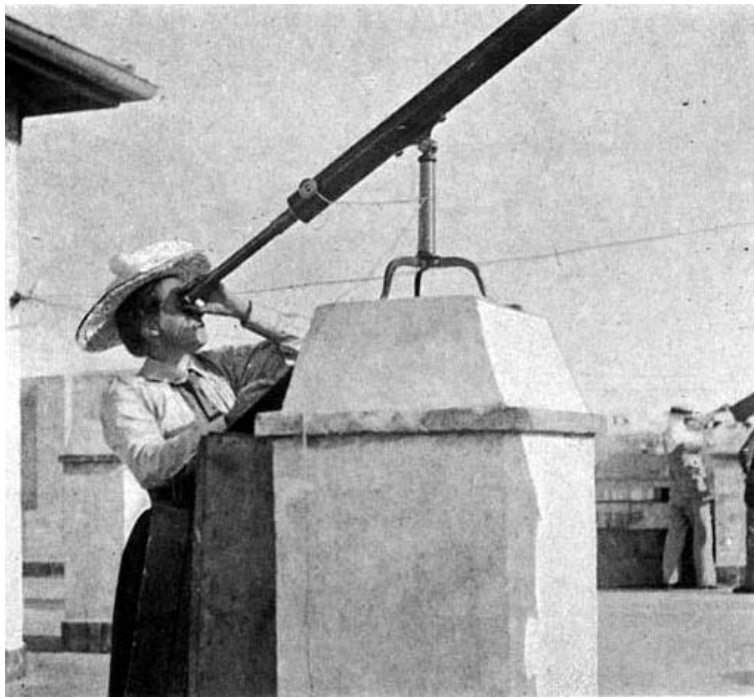

**Figure 22: Observer using a chimney as a telescope support on the roof of the Hotel de la Régence, Algiers** (82)

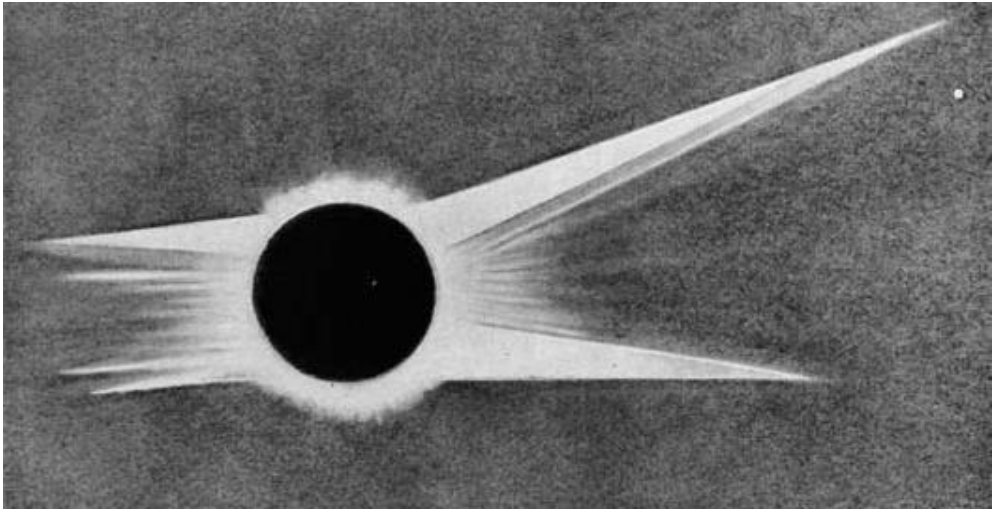

**Figure 23: Drawing of the total solar eclipse of 1900 as observed from Algiers**
(82)

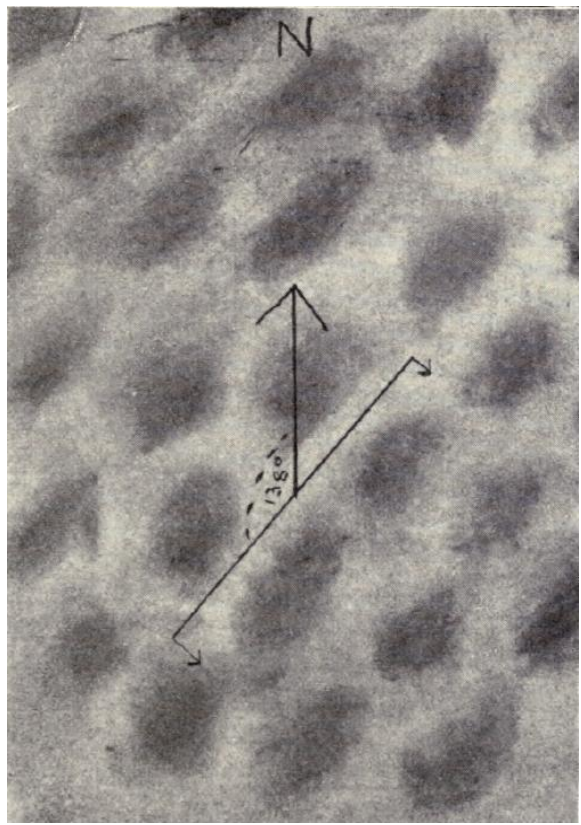

**Figure 24: Drawing of the shadow bands**

Drawn by Ruth Mary Brook (82). The line at 138⁰ from North indicates the orientation of the shadow bands and the arrows at the end of the lines indicates the direction of travel

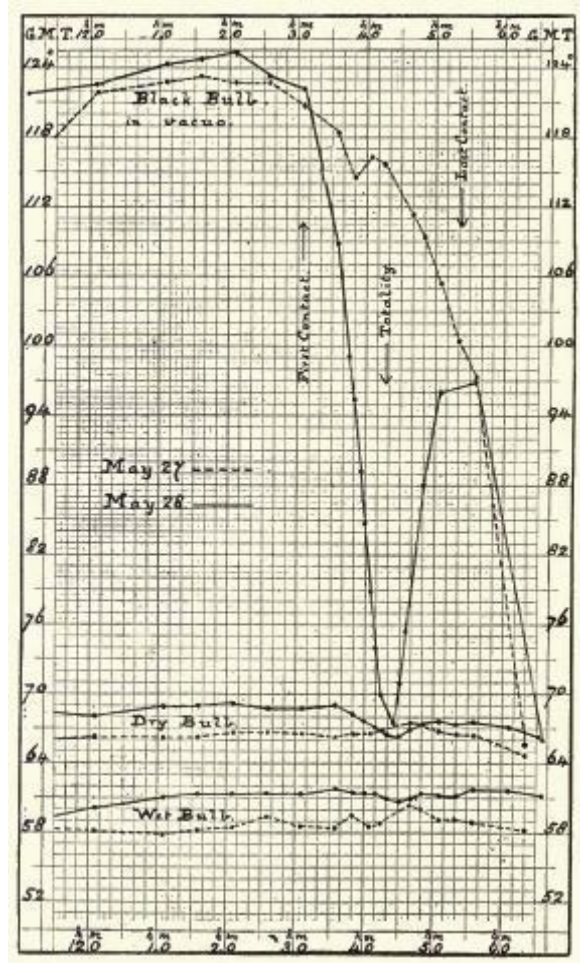

**Figure 25: Temperature observations**

Made by C.L. Brook on 27[th] and 28[th] May 1900 (the latter being the day of the eclipse) at the Hôtel de la Régence, Algiers